\documentclass[twocolumn,showpacs,preprintnumbers,amsmath,amssymb,superscriptaddress,showkeys]{revtex4}
\usepackage{amsmath}
\usepackage{amsfonts}
\usepackage{amssymb}
\usepackage{graphicx}
\usepackage{natbib}
\usepackage[dvipdfm]{hyperref}

\begin{document}
\title{Negativity of quantumness and non-Markovianity in a qubit coupled to a thermal Ising spin bath system}
\author{Zheng-Da Hu}\email{huyuanda1112@jiangnan.edu.cn}
\affiliation{School of Science, Jiangnan University, Wuxi 214122, China}
\author{Yixin Zhang}
\affiliation{School of Science, Jiangnan University, Wuxi 214122, China}
\author{Ye-Qi Zhang}
\affiliation{Department of Mathematics and Physics, North China Electric Power University, Beijing 102206, China}
\begin{abstract}
We propose a scheme to characterize the non-Markovian dynamics and quantify the non-Markovianity via the non-classicality measured by the negativity of quantumness. By considering a qubit in contact with a critical Ising spin bath and introducing an ancilla, we show that revivals of negativity of quantumness indicate the non-Markovian dynamics. Furthermore, a normalized measure of non-Markovianity based on the negativity of quantumness is introduced and the influences of bath criticality, bath temperature and bath size on the non-Markovianity are discussed. It is shown that, at the critical point, the decay of non-Markovianity versus the size of spin bath is fastest and the non-Markovianity is exactly zero only in the thermodynamic limit. Besides, non-trivial behaviours of negativity of quantumness such as sudden change, double sudden changes and keeping constant are found for different relations between parameters of the initial state. Finally, how the non-classicality of the system is affected by a series of bang-bang pulses is also examined.
\end{abstract}
\maketitle

\section{Introduction}
Decoherence may occur everywhere and cause great trouble in implementing quantum tasks.
This is due to the unavoidable coupling between any realistic quantum system and its environment,
which may lead fast destruction of quantum superposition~\cite{RevModPhys.75.715,oqs}.
To understand decoherence, the environment of a quantum system is paradigmatically modeled as a many-body system,
such as a set of harmonic oscillators~\cite{qds} or spins~\cite{spinbath}.
Much attention has been paid to the spin baths,
since quantum spin systems play an important role in quantum information processing and condensed matter physics~\cite{PhysRevLett.96.140604,PhysRevA.75.032109,Hu20123011,PhysRevA.77.052112,PhysRevA.79.012305,PhysRevLett.109.185701}.
Especially, decoherence quantified by the decay of Loschmidt echo~\cite{PhysRevLett.91.210403} can be greatly enhanced by a
critical spin bath~\cite{PhysRevLett.96.140604,PhysRevA.79.012305}.
Meanwhile, several schemes have been proposed to deal with decoherence, including decoherence free subspace~\cite{lidar_decoherence-free_1998},
quantum Zeno effect~\cite{maniscalco_protecting_2008}, and dynamical decoupling~\cite{viola_dynamical_1999}, etc.

Recently, it has been realized that entanglement represents only a special kind of nonclassical correlation.
Even unentangled (separable) states show some nonclassical phenomena,
which can be captured by a new kind of nonclassical correlation termed as quantum discord~\cite{RevModPhys.84.1655}.
The original quantum discord is defined by
the difference between the quantum mutual information and the classical correlation~\cite{henderson_classical_2001,ollivier_quantum_2001}.
Other measures of discord such as relative entropy of discord~\cite{PhysRevLett.104.080501}, geometric discord~\cite{PhysRevLett.105.190502}, trace-distance discord~\cite{PhysRevA.87.064101,PhysRevA.87.042115} have also been proposed
based on the idea that the desired correlation is the distance from a given state to the closest state without the desired property. For pure
states, discord is equivalent to entanglement, while for general mixed states, it is more robust against decoherence than entanglement~\cite{werlang_robustness_2009}. Rather than suddenly vanishing of entanglement in a finite time (entanglement sudden death~\cite{yu_sudden_2009}) under decoherence,
the quantum discord vanishes asymptotically and may exhibit the phenomenon of sudden change~\cite{PhysRevA.80.044102} or sudden transition~\cite{PhysRevLett.104.200401}. Very recently, a discord-like quantifier termed as negativity of quantumness has also been proposed~\cite{PhysRevLett.106.220403,PhysRevA.88.012117} and experimentally reported~\cite{PhysRevLett.110.140501,PhysRevLett.111.250401}.
Negativity of quantumness is the minimum negativity created between the system and the apparatus which performs local measurements on subsystems and thus quantifies the degree of non-classicality on the measured subsystems determined by
which and how many subsystems are measured~\cite{PhysRevA.88.012117}.

On the other hand, a precise description of the open dynamics process is also desirable.
When the environment is infinitely sized and weakly coupled to the quantum system,
the reduced system dynamics under Born-Markov approximation
can be treated as a Markovian process and a master equation of Lindblad form can be derived~\cite{JMP.17.821,CMP.48.119}.
The Markovian process should be memoryless, and revival dynamics is usually referred to a signature of non-Markovian effect~\cite{PhysRevLett.101.150402,PhysRevLett.103.210401,PhysRevLett.105.050403}

However, the non-Markovianity, which is a measure of the degree of non-Markovian effect in open systems, is usually difficult to calculate~\cite{PhysRevLett.103.210401,PhysRevLett.105.050403} due to the optimization over all pairs of initial states and the accumulation
of all information back flowed. In this paper, we propose a scheme to characterize the non-Markovian dynamics and explore the non-Markovianity via the non-classicality measured by the negativity of quantumness, which may be more convenient to calculate. By considering a qubit coupled to a thermal Ising spin bath and introducing an ancilla, we study the dynamics of non-classicality of the qubit system. It is shown that revivals of the negativity of quantumnes can be treated as a signature of non-Markovian dynamics. Moreover, we introduce a measure of the non-Markovianity based on the negativity of quantumness and investigate the influences of bath criticality, bath temperature and bath size on the non-Markovianity. It is observed that the non-Markovianity converges to zero very sensitively at the critical point as the bath size is enlarged. Besides, it is also interesting to explore the behaviour of the negativity of quantumness to compare with that of other discord-like quantifiers where the phenomenon of sudden change~\cite{PhysRevA.80.044102} or sudden transition~\cite{PhysRevLett.104.200401} may occur. We find non-trivial behaviours of negativity of quantumness such as sudden change, double sudden changes and keeping constant for different relations between parameters of the initial state.
It is also shown that the negativity of quantumness is greatly destroyed by the critical spin bath. Therefore, we use the scheme of dynamical decoupling (bang-bang control) to protect the quantumness of the system.

This paper is organized as follows. In Sec.~\ref{sec:sec2}, we introduce the model as a qubit coupled to a thermal Ising spin bath and derive the reduced dynamics analytically. By introducing an ancilla, the non-Markovianity based on the negativity of quantumness is discussed. In Sec.~\ref{sec:sec3}, we proceed to investigate dynamical properties of negativity of quantumness and how it is affected by the criticality of the spin bath. Sec.~\ref{sec:sec4} is devoted to studying how the non-classicality of the qubit system is affected by a series of bang-bang pulses. A summary is given at last in Sec.~\ref{sec:sec5}.

\section{Non-Markovianity of an open quantum system from a thermal spin bath}\label{sec:sec2}
\begin{figure}[!htb]
\centering
\includegraphics[width=6.5cm]{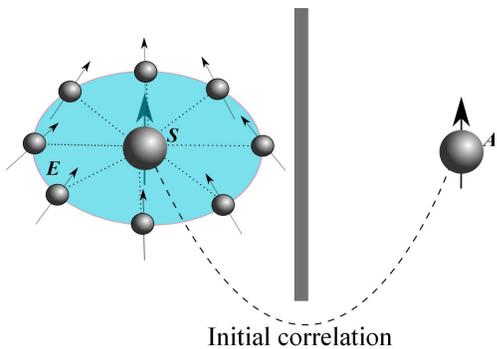}
\caption{(Color online) The qubit system $S$ is disturbed by its environment $E$ which is modeled by a quantum Ising bath. To describe the non-Markovian dynamics via the non-classicality (negativity of quantumness), an auxiliary qubit termed as the ancilla $A$ is introduced to share certain amount of quantum correlation with the system $S$.}
\label{fig:fig1}
\end{figure}
We consider an open system represented by a qubit (denoted as $S$) coupled to a thermal Ising spin bath
(denoted as $E$) in a dephasing way~\cite{PhysRevLett.96.140604,PhysRevA.75.032109,Hu20123011,PhysRevA.77.052112}, which is sketched in the left part of Fig.~\ref{fig:fig1}. The Hamiltonian is given by $H_{0}=H_{S}+H_{E}+H_{SE}$ with
\begin{eqnarray}
H_{S} & = & f\sigma_{S}^{z},\nonumber\\
H_{E} & = & J\sum_{i=1}^{N}(\sigma_{i,E}^{x}\sigma_{i+1,E}^{x}+h\sigma_{i,E}^{z}),\\
H_{SE} & = & \varepsilon\frac{(I_{S}-\sigma_{S}^{z})}{2}\otimes\sum_{j=1}^{N}\sigma_{j,E}^{z},\nonumber
\end{eqnarray}
where $I_{S}$ is the identity operator of the system $S$, $\sigma_{S}^{\alpha}$ and $\sigma_{i,E}^{\alpha}$ are the standard Pauli matrixes
(in the basis $\{\vert\uparrow\rangle,\vert\downarrow\rangle\}$) for the system $S$ and the $i$th spin of the bath $E$, respectively.
The parameter $f$ in $H_{S}$ is related to the transition frequency and $h$ in $H_{E}$ measures the strength of external transverse field.
The constant $J$ characterizes the coupling strength between nearest-neighbor spins of the bath
and $\varepsilon$ denotes the coupling between the qubit and its bath.
The reduced dynamics of the system $S$ is exactly solvable as long as one knows the decoherence factor
\begin{equation}
\mathcal{F}(t)=\textrm{Tr}(\rho_{E}\mathrm{e}^{\mathrm{i}H_{E\uparrow}t}\mathrm{e}%
^{-\mathrm{i}H_{E\downarrow}t}),
\end{equation}
with $H_{E\uparrow }=f+H_{E}(h )$, $H_{E\downarrow }=-f+H_{E}(\tilde{h})$ and $\tilde{h}=h +\varepsilon/J$.
Here, we assume the spin bath $E$ is in its thermal state, $\rho_{E}=\exp(-\beta H_{E})/Z_{E}$,
where $Z_{E}=\mathrm{Tr}[\exp(-\beta H_{E})]$ is the partition function
and $\beta=1/(\kappa_{B}T)$ with $\kappa_{B}$ the Boltzmann constant and $T$ the temperature.
Then the explicit form of the decoherence factor can be expressed as~\cite{PhysRevLett.96.140604,PhysRevA.75.032109,Hu20123011}
\begin{eqnarray}
\mathcal{F}(t)&=&\mathrm{e}^{\mathrm{i}2ft}\prod_{k>0}\frac{1}{z_{k}}\{\exp[2J\beta\Lambda_{k}(h)-\mathrm{i}g_{k}]\nonumber\\
&\quad&\times[\cos\tilde{g_{k}}+\mathrm{i}\sin\tilde{g_{k}}\cos(2\alpha_{k})]\nonumber\\
&\quad&+\exp[-2J\beta\Lambda_{k}(h)+\mathrm{i}g_{k}]\nonumber\\
&\quad&\times[\cos\tilde{g_{k}}-\mathrm{i}\sin\tilde{g_{k}}\cos(2\alpha_{k})]\},
\end{eqnarray}
with $z_{k}=2\cosh[2J\beta\Lambda_{k}(h)]$, $g_{k}=2J\Lambda_{k}(h)t$, and $\alpha_{k}=(\tilde{\theta}_{k}-\theta_{k})/2$,
where
\begin{eqnarray}
\Lambda_{k}(h) & = & \sqrt{(\cos k+h)^{2}+\sin^{2}k},\nonumber\\
\cos\theta_{k} & = & (\cos k+h)/\Lambda_{k}(h),\\
\sin\theta_{k} & = & \sin k/\Lambda_{k}(h),\nonumber
\end{eqnarray}
and $k=\frac{\pi}{N},\frac{3\pi}{N},...,\frac{(N-1)\pi}{N}$.
Here, it is worth noting that $\tilde{g_{k}}$ and $\tilde{\theta}_{k}$ have the same
forms as $g_{k}$ and $\theta_{k}$ respectively, simply by replacing
$h$ with $\tilde{h}=h+\varepsilon/J$.

The information exchange between the qubit and the spin bath can be
quantified by the Loschmidt echo~\cite{PhysRevLett.91.210403}. The Loschmit echo that introduced
in NMR experiments to describe the hypersensitivity of the time
evolution to the environmental effects  is defined by
\begin{equation}\label{Le}
\mathcal{L}(t)=\vert\mathcal{F}(t)\vert^{2}.
\end{equation}
A simple relationship between the Loschmidt echo and the non-Markovianity has been found~\cite{PhysRevA.85.060101}, i.e.,
a monotonous decay of $\mathcal{L}(t)$ is a signature of Markovian dynamics,
while a increasing of $\mathcal{L}(t)$ at any time instant is a direct signature of non-Markovian dynamics (backflow of information).
Here, we proposed an alternative manner via quantum correlation to describe the non-Markovianity motivated by~\cite{PhysRevLett.105.050403}.
First, as is shown in the right part of Fig.~\ref{fig:fig1}, an ancilla $A$ (another qubit) is introduced ,
which does not interact with $SE$ but initially shares certain amount of quantum correlation with the qubits $S$.
In this sense, the total system $SEA$ and the subsystem $SE$ both evolve unitarily and the ancilla $A$ does not evolve.
Here, we use the quantum correlation called negativity of quantumness~\cite{PhysRevA.88.012117} as a measure of non-Markovianity, which is contractive under Markovian channels and may be a suitable quantity for revealing the environmental memory effect.

First, we briefly outline some concepts for the negativity of quantumness.
The negativity of quantumness is a measure of non-classicality which recently is theoretically proposed~\cite{PhysRevLett.106.220403,PhysRevA.88.012117} and experimentally reported~\cite{PhysRevLett.110.140501}.
It corresponds to the minimum negativity created between the system and the apparatus which performs local measurements on subsystems.
The negativity of quantumness thus quantifies the degree of non-classicality on the measured subsystems determined by
which and how many subsystems are measured~\cite{PhysRevA.88.012117}.
When the measured subsystem is a qubit ($S$ here), the negativity
of quantumness measuring the non-classicality of subsystem $S$ can be expressed as
\begin{equation}
\mathcal{Q}_{S}=\vert\vert\rho_{SA}-\chi_{S}\vert\vert_{1},
\end{equation}
where $\vert\vert X\vert\vert_{1}=\mathrm{Tr}(\sqrt{X^{\dagger}X})$
is the trace norm and a normalization factor $2$ has been multiplied.
In this case, the negativity of quantumness has a good geometric interpretation
such that it is also termed as the one-norm geometric quantum discord~\cite{PhysRevA.87.064101,PhysRevA.87.042115}.

Then, by considering the optimal initial system-ancilla state~\cite{PhysRevLett.105.050403}, i.e., an initial maximally entangled state,
the non-Markovianity can be defined in a similar way to Ref.~\cite{PhysRevLett.103.210401,PhysRevLett.105.050403} as
\begin{eqnarray}
\mathcal{N}_{\mathcal{Q}} & = &\int_{\dot{\mathcal{Q}}_{S}>0}\dot{\mathcal{Q}}_{S}\mathrm{\, d}t= \sum_{i}(\sqrt{\mathcal{L}(t_{i}^{\max})}-\sqrt{\mathcal{L}(t_{i}^{\min})}),
\end{eqnarray}
where $\mathcal{L}(t)$ is the Loschmidt echo given by Eq.~(\ref{Le}),
and $t_{i}^{\max}$ ($t_{i}^{\min}$) is the time point of the $i$th local maximum (minimum) of $\mathcal{L}(t)$ during the time $t\in(0,\,\infty)$.
It should be noted that the non-Markovianity may diverge~\cite{PhysRevLett.103.210401} if non-Markovian revivals are infinite. To avoid the
divergence, a normalized version of non-Markovianity~\cite{PhysRevLett.105.050403} can be introduced as $\mathcal{I_{Q}=N_{Q}}/(\mathcal{N_{Q}}+1)$,
which is a monotonically increasing function of $\mathcal{N_{Q}}$ such that $\mathcal{I_{Q}}=0$ for $\mathcal{N_{Q}}=0$ and $\mathcal{I_{Q}}=1$
for $\mathcal{N_{Q}}=\infty$. Here, we define an alternative measure of normalized non-Markovianity as
\begin{equation}
\mathcal{N}=\max_{\{t_{j}\geq t_{i}\}}\frac{\mathcal{Q}_{S}(t_{j}^{\max})-\mathcal{Q}_{S}(t_{i}^{\min})}{\mathcal{Q}_{S}(0)-\mathcal{Q}_{S}(t_{i}^{\min})},
\end{equation}
which calculates the maximum reversal (backflow) of quantumness in ratio to the total amount of quantumness lost previously and thus quantifies the memory ability of the environment. It is clear that $\mathcal{N}$ is bounded from
zero to unit, such that $\mathcal{N}=0$ when the system continuously loses its quantumness, and $\mathcal{N}=1$ when the quantumness of system is fully recovered to its initial value.

\begin{figure}[!htb]
\centering
\includegraphics[width=6.5cm]{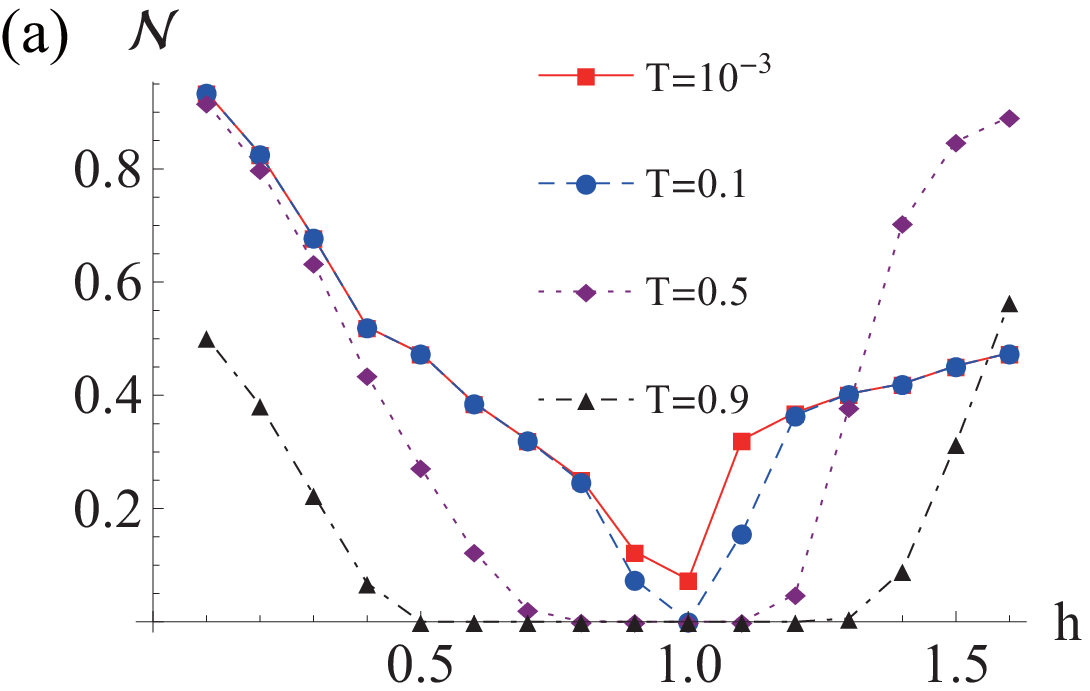}
\includegraphics[width=6.5cm]{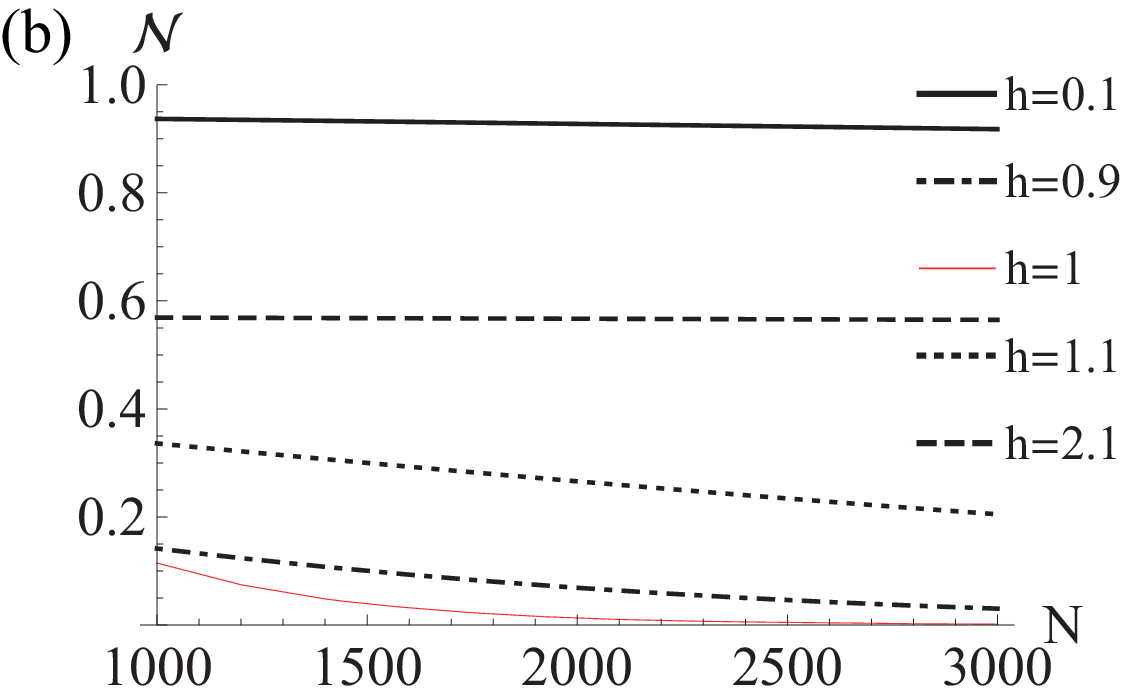}
\caption{(Color online) (a) Non-Markovianity $\mathcal{N}$ as a function of bath field $h$ for the spin bath of $N=1200$ at different temperatures. (b) Non-Markovianity $\mathcal{N}$ versus the spin bath $N$ with different strength of the bath field at temperature $T=10^{-3}$. Other parameters in the two panels are chosen as $\kappa_{B}=J=1$ and $\varepsilon=0.05$.}
\label{fig:fig2}
\end{figure}

In order to explore how the non-Markovianity $\mathcal{N}$ is influenced by the properties of the spin bath such as criticality, temperature and size, we first plot the non-Markovianity $\mathcal{N}$ as a function of the transverse field $h$ of the Ising bath with different temperatures $T$ in Fig.~\ref{fig:fig2}(a). It is clearly seen from Fig.~\ref{fig:fig2}(a) that when the temperature $T$ is low ($T=10^{-3}$ and $T=0.1$) the non-Markovianity decreases sharply near the critical point $h_{\mathrm{c}}=1$. As the temperature increases, the non-Markovianity near the critical point decreases further and will go to zero at higher temperature. Generally, the non-Markovianity at non-critical points decrease and will also drop to near zero when the bath temperature is enhanced. However, there exists a region ($h>1.3$ shown in Fig.~\ref{fig:fig2}(a)) where the non-Markovianity even increases versus temperature at first (comparing the value at $T=0.5$ to that at $T=0.1$) and decreases as the temperature is further enhanced (comparing the value at $T=0.9$ to that at $T=0.5$). It is also noted that the non-Markovianity with finite sized spin bath may not be zero at the critical point, which means that the dynamics is not exactly Markovian and  certain revivals of quantumness is still allowed. In spite of the finite size effect, the non-Markovianity in the critical region is the most sensitive to the growth of bath size. To show this, the non-Markovianity $\mathcal{N}$ with different strengths of bath field versus the bath size $N$ is plotted in Fig.~\ref{fig:fig2}(b). We observe that the non-Markovianity decreases in a polynomial manner at non-critical points. By contrast, it decays in an near exponential manner at the critical point (shown by the red curve), indicating the sensitivity of non-Markovianity to the finite bath effect near the critical point and purely Markovian dynamics in the thermodynamic limit.

\section{Dynamics of negativity of quantumness from thermal spin bath}\label{sec:sec3}

\begin{figure}[!htb]
\centering
\includegraphics[angle=0,width=6cm]{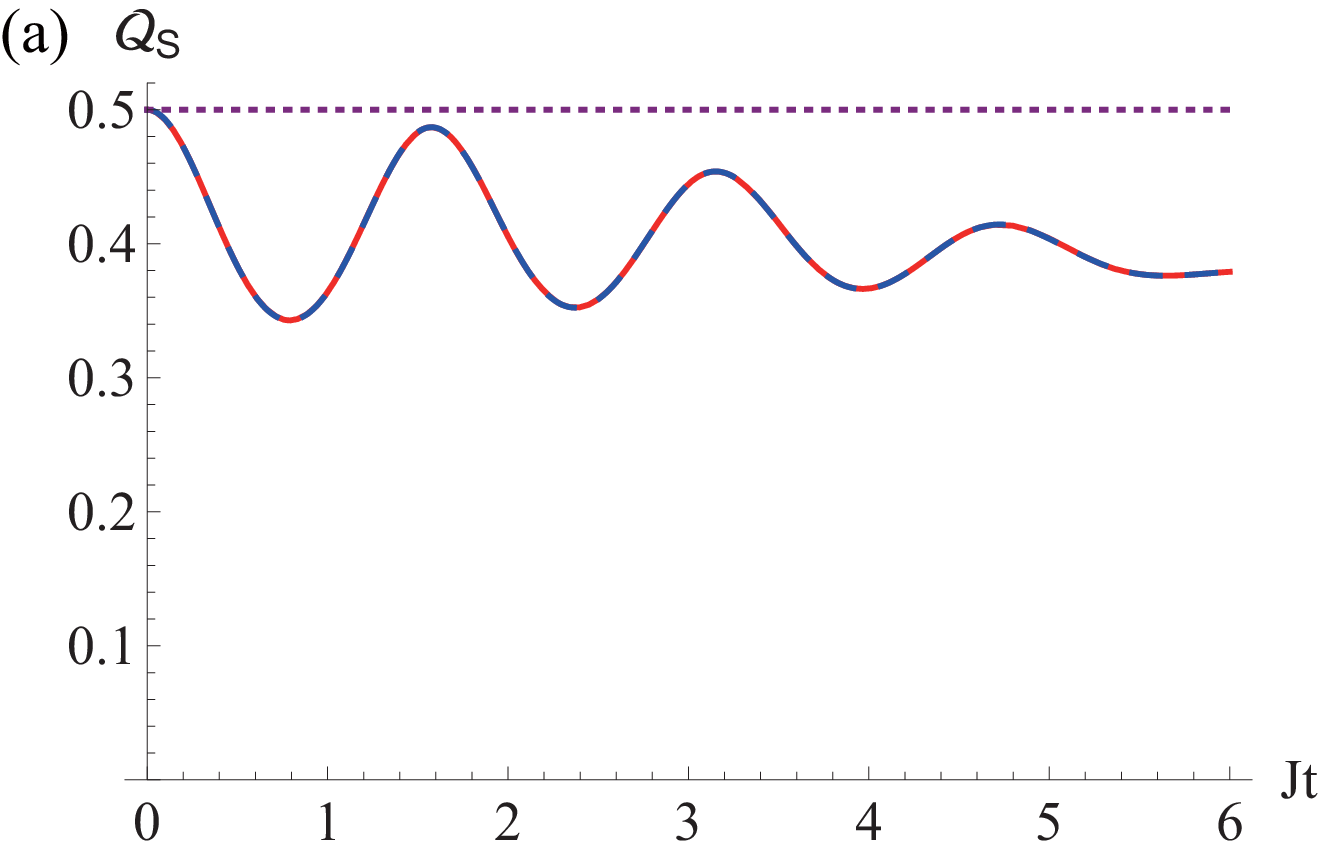}
\includegraphics[angle=0,width=6cm]{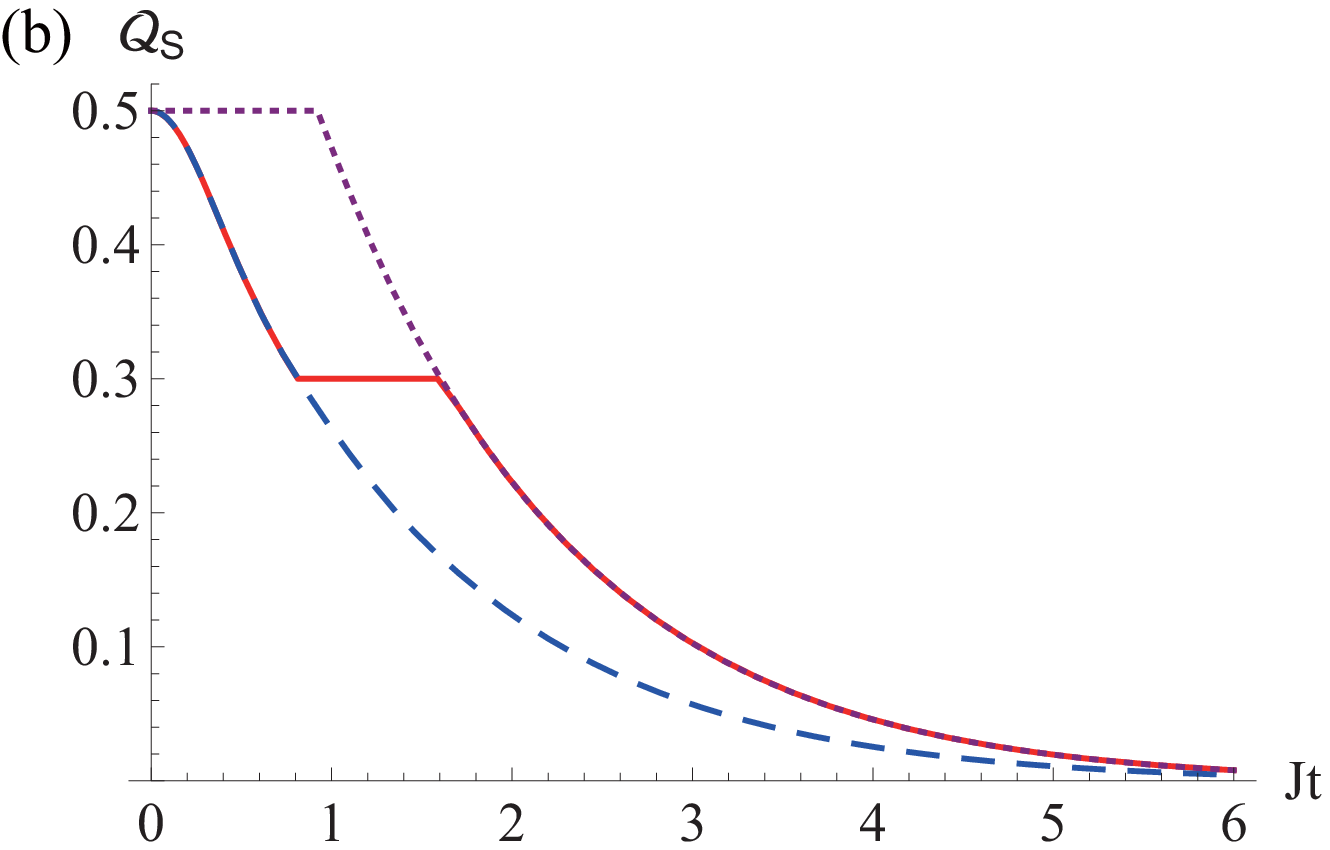}
\includegraphics[angle=0,width=6cm]{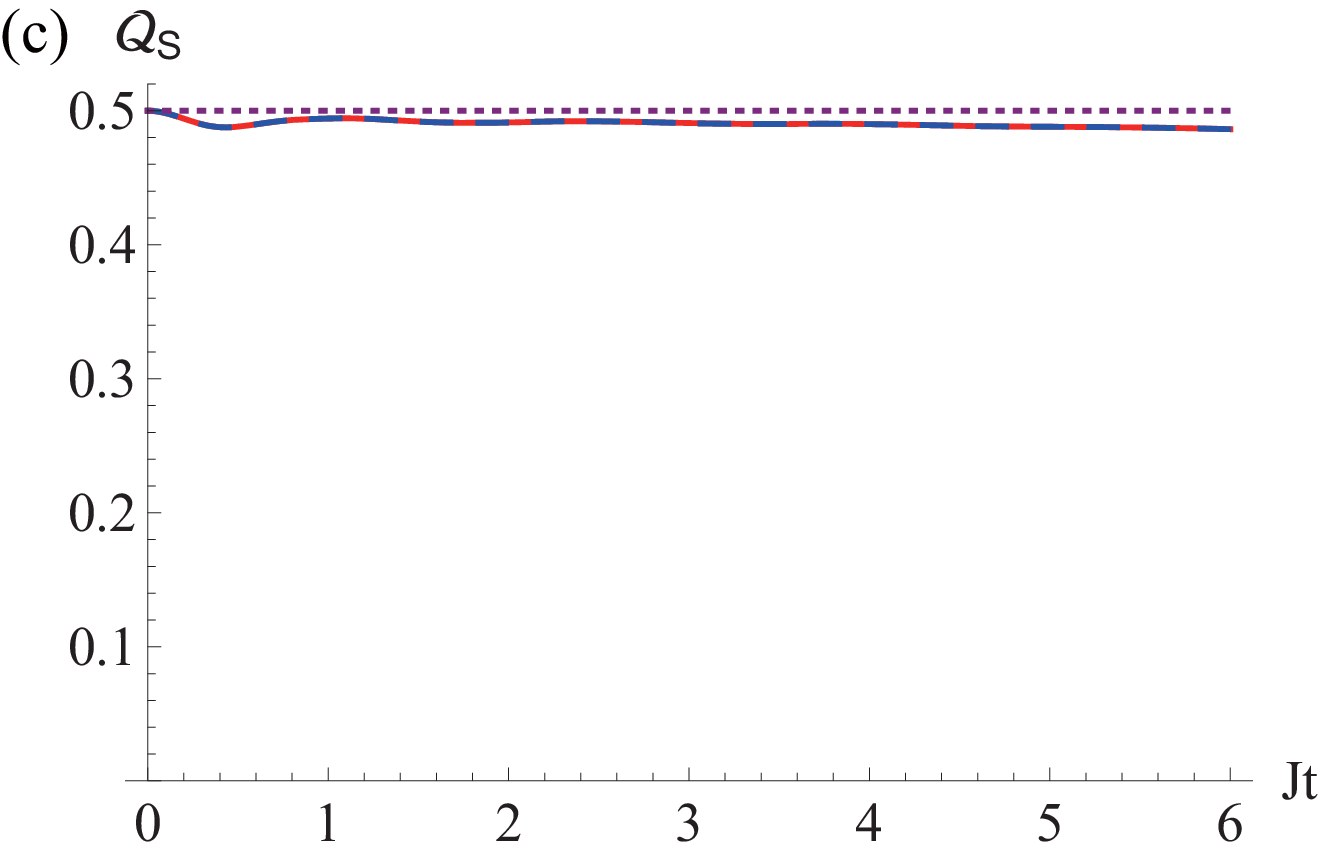}
\caption{(Color online) Dynamics of negativity of quantumness $\mathcal{Q}_{S}$ as a function of rescaled time $Jt$ in different regions with different relations between the coefficients $c_{i}$ of the initial state. Non-trivial variations of $\mathcal{Q}_{S}$ are presented in three regions (a) $h\ll1$ ($h=0.1$), (b) $h\simeq1$ ($h=1$) and (c) $h\gg1$ ($h=2$). The blue-dashed curers ($c_1=0.5$, $c_2=0.3$ and $c_3=0.9$), purple-dotted curves ($c_1=0.9$, $c_2=0.3$ and $c_3=0.5$) and red-solid curves ($c_1=0.9$, $c_2=0.5$ and $c_3=0.3$) correspond to cases (i) $\vert c_{3}\vert>\vert c_{1}\vert,\vert c_{2}\vert$, (ii) $\vert c_{1}\vert>\vert c_{3}\vert>\vert c_{2}\vert$ and (iii) $\vert c_{1}\vert,\vert c_{2}\vert>\vert c_{3}\vert$, respectively. In the three panels, other parameters are chosen as $N=1200$, $T=0.5$, $\varepsilon=0.05$, $J=\kappa_{B}=1$.}
\label{fig:fig3}
\end{figure}

In this section, we proceed to investigate dynamical properties of negativity of quantumness
$\mathcal{Q}_{S}(t)$ and how it is affected by the criticality of the spin bath. It is interesting to explore the dynamics of the negativity of quantumness and ensure that whether negativity of quantumness behaves as other discord-like quantifiers where the phenomenon of sudden change~\cite{PhysRevA.80.044102} or sudden transition~\cite{PhysRevLett.104.200401} may occur.

We assume the initial state of qubits $SA$ is prepared in a Bell diagonal state as~\cite{PhysRevLett.104.200401}
\begin{equation}\label{Bds}
\rho_{SA}(0)=\frac{1}{4}(I_{SA}+\sum_{i=1}^{3}c_{i}\sigma_{S}^{i}\otimes\sigma_{A}^{i}),
\end{equation}
with $c_{i}$ and $\sigma^{i}$ denoting real coefficients and the standard Pauli matrixes, and the spin bath has been in its thermal equilibrium state. Then the reduced state of the two-qubit subsystem $SA$ at time $t$ is given by
\begin{equation}\label{Bdst}
\rho_{SA}(t)=\left(\begin{array}{cccc}
a &  &  & w^{\ast}\\
 & b & z\\
 & z & b\\
w &  &  & a
\end{array}\right),
\end{equation}
where $a=(1+c_{3})/4$, $b=(1-c_{3})/4$, $z=(c_{1}+c_{2})\left|\mathcal{F}(t)\right|/4$ and $w=(c_{1}-c_{2})\mathcal{F}(t)/4$. The state $\rho_{SA}(t)$ above is not a Bell diagonal state. However, under a local unitary operation $V=\exp(-\mathrm{i}\phi\sigma^{z}_{S}/2)$$\otimes\exp(-\mathrm{i}\phi\sigma^{z}_{A}/2)$, where $\phi$ is the argument of the decoherence factor $\mathcal{F}(t)$, such that
\begin{equation}
\tilde{\rho}_{SA}(t)= V^{\dagger}\rho_{SA}(t)V= \frac{1}{4}(I_{SA}+\sum_{i=1}^{3}c_{i}(t)\sigma_{S}^{i}\otimes\sigma_{A}^{i}),
\end{equation}
with $c_{1}(t)=c_{1}\left|\mathcal{F}(t)\right|$, $c_{2}(t)=c_{2}\left|\mathcal{F}(t)\right|$
and $c_{3}(t)=c_{3}$, it is again a Bell diagonal state. Since local unitary operations do not change quantum correlations, the negativity
of quantumness for the state (\ref{Bdst}) is then given by~\cite{PhysRevA.88.012117}
\begin{equation}\label{nq}
\mathcal{Q}_{S}(\rho_{SA}(t))=\mathrm{int}\{\vert c_{1}(t)\vert,\vert c_{2}(t)\vert,\vert c_{3}(t)\vert\}.
\end{equation}
It can be seen from Eq.~(\ref{nq}) that the negativity of quantumness depends on the relations between the coefficients $c_{i}$ of the state.

(i) If $\vert c_{3}\vert>\vert c_{1}\vert,\vert c_{2}\vert$, the negativity of quantumness reads as $\mathcal{Q}_{S}(t)=c_{0}\sqrt{L(t)}$ with $c_{0}=\min\{\vert c_{1}\vert,\vert c_{2}\vert\}$, which is exactly the square root of Loschmidt echo multiplied by a constant factor. In this case, any revival of negativity of quantumness is a signature of non-Markovian effect. It can be seen from the blue-dashed curves in Fig.~\ref{fig:fig3}(a-c) that the decay of $\mathcal{Q}_{S}(t)$ can be explicitly revivable, be significantly enhanced, and be greatly suppressed, in the regions $h\ll1$, $h\simeq1$ and $h\gg1$ respectively.

(ii) If $\vert c_{1}\vert>\vert c_{3}\vert>\vert c_{2}\vert$ or $\vert c_{2}\vert>\vert c_{3}\vert>\vert c_{1}\vert$, then in the regions $h\ll1$ or $h\gg1$, $\mathcal{Q}_{S}(t)$ equals to $c_{3}$, which is a constant and shown by the purple-dotted curve in Fig.~\ref{fig:fig3}(a) and (c). This means that the decoherence is not such strong that disturbs the negativity of quantumness in this two non-critical regions. However, in the critical region $h\simeq1$, the decoherence is enhanced rapidly and strong enough to destroy the negativity of quantumness and then the phenomenon of sudden change~\cite{PhysRevA.80.044102} of quantum correlation occurs. After keeping constant for a while, the negativity of quantumness starts to decay suddenly (sudden change), which is shown by the purple-dotted curve in Fig.~\ref{fig:fig3}(b).

(iii) If $\vert c_{1}\vert,\vert c_{2}\vert>\vert c_{3}\vert$,
the dynamics of $\mathcal{Q}_{S}(t)$ is highly dependent on the criticality of the spin bath.
Different characteristic behaviours exist respect to different regions of external field $h$ of the spin bath. As is displayed by the red-solid curves in Fig.~\ref{fig:fig3}(a-c), the dynamics of $\mathcal{Q}_{S}(t)$ can exhibit explicit non-Markvian revivals, double sudden changes and immunity to decoherence decay in the regions $h\ll1$, $h\simeq1$ and $h\gg1$ respectively. It is worth noting that this type behaviour of environment-induced double sudden changes has been experimentally reported very recently~\cite{PhysRevLett.111.250401}.

Overall, the dynamics of negativity of quantumness $\mathcal{Q}_{S}(t)$ depends on both the initial state parameters $c_{i}$ and the external field $h$. Some interesting results can be drawn until now. First,  the negativity of quantumness is not disturbed in the regions $h\ll1$ or $h\gg1$ for certain values of $c_{i}$. Second, the decoherence is qualitatively enhanced by the quantum phase transition of spin bath leading to phenomena of fast decay, sudden change and double sudden changes for different collocations of $c_{i}$. Finally, the negativity of quantumness can also be well preserved by enhancing the external field $h$ to far away from the critical point.

\section{Dynamical decoupling from thermal spin bath}\label{sec:sec4}
\begin{figure}[!htb]
\centering
\includegraphics[angle=0,width=7cm]{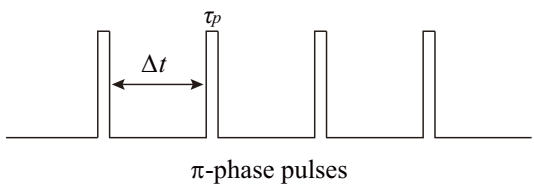}
\caption{Bang-bang control scheme: $\pi$-phase pulses of constant amplitude $\pi/(2\tau_{p})$ with each duration $\tau_{p}$ are applied periodically to the system $S$ which lead to nearly instant spin flips. Here $\Delta t$ denotes the interval of adjacent pulses.}
\label{fig:fig4}
\end{figure}

\begin{figure}[!htb]
\centering
\includegraphics[angle=0,width=6cm]{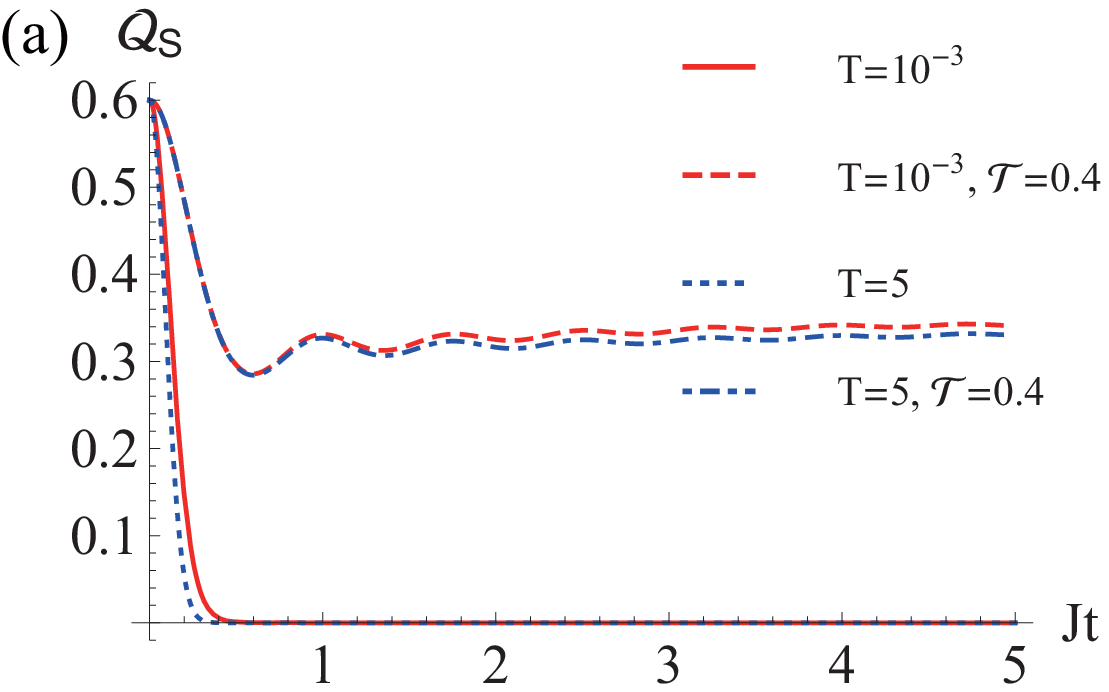}
\includegraphics[angle=0,width=6cm]{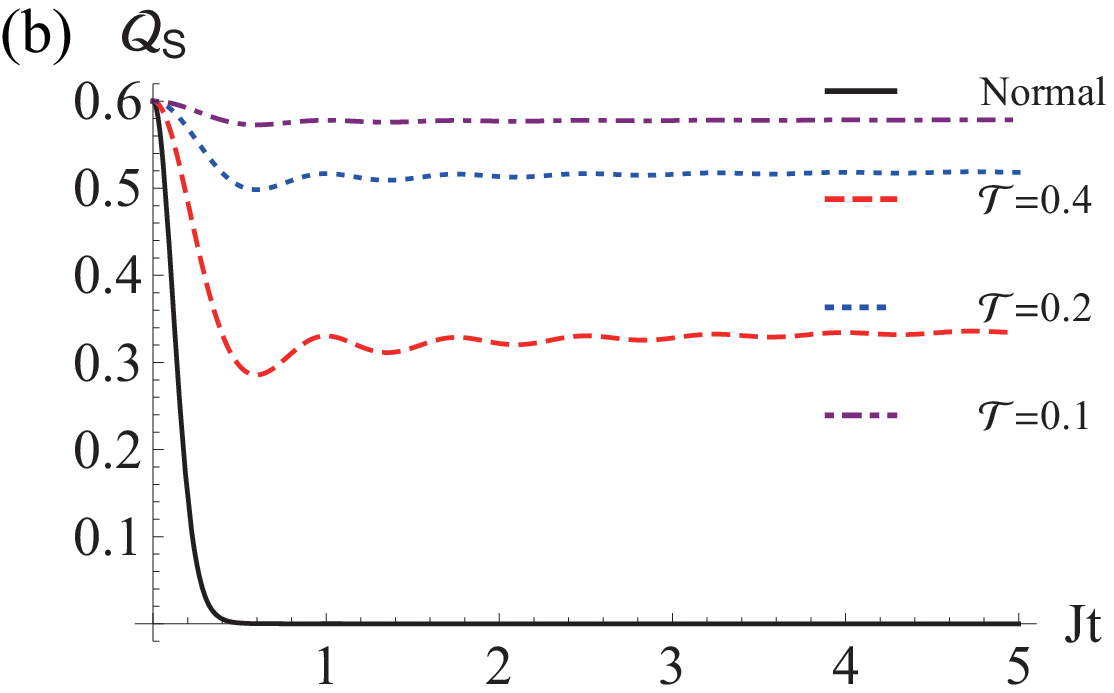}
\caption{(Color online) (a) Negativity of quantumness $\mathcal{Q}_{N}^{S}$ as a function of rescaled time $Jt$ with (dashed and dot-dashed curves) and without (solid and dotted curves) pulses control at near zero temperature (red curves $T=10^{-3}$) and at finite bath temperature (blue curves $T=5$). (b) Negativity of quantumness as a function of rescaled time $Jt$ under pulses control with different frequencies (dashed, dotted and dot-dashed curves corresponding to $\mathcal{T}=0.4,0.2,0.1$ respectively) and without pulses control (solid curve) for specific bath temperature $T=0.5$. In the two panels, other parameters are chosen as $N=1200$, $\varepsilon=0.25$, $h=J=\kappa_{B}=1$, $c_{1}=c_{2}=0.6$ and $c_{3}=0.8$.}
\label{fig:fig5}
\end{figure}

In the section above, we have shown that the negativity of quantumness can be seriously destroyed by the critical spin bath (see Fig.~\ref{fig:fig3}b). It is then necessary to find efficient ways to protect the quantumness of a quantum system.
In this section, we utilize a dynamical decoupling technique to modulate the dynamics of $\mathcal{Q}_{S}(t)$.
The dynamical decoupling techniques have been shown to effectively prevent quantum systems from decoherence~\cite{PhysRevA.77.052112,viola_dynamical_1999}.
Here we apply a sequence of bang-bang pulses ($\pi$-pulses control sketched in Fig.~\ref{fig:fig4}) to the qubit $S$, where each pulse of duration $\tau_{p}$ causes an instant spin flip after every normal evolution interval $\Delta t$. The Hamiltonian under pulses control is then given by
\begin{equation}
H=H_{0}+H_{p},
\end{equation}
with
\begin{equation}
H_{p}=\frac{\pi}{2\tau_{p}}\sum_{n=0}^{\infty}\theta(t-\Delta t-t_{n})\theta(t_{n}+\Delta t+\tau_{p}-t)\sigma_{S}^{x},
\end{equation}
where $\theta(x)$ is a step function of $x$ and $t_{n}=n(\Delta t+\tau_{p})$.
Treating twice pulse actions as a complete cycle, the dynamical evolution
of a complete cycle can be expressed as
\begin{equation}
U_{c}(\mathcal{T})=U(t_{2(n+1)},t_{2n})=U(\tau_{p})U_{0}(\Delta t)U(\tau_{p})U_{0}(\Delta t),
\end{equation}
with $U_{0}(\Delta t)=\exp(-\mathrm{i}H_{0}\Delta t)$ and $U(\tau_{p})=\exp(-\mathrm{i}H\tau_{p})$
describing the evolution operators during the normal evolution and the action
of pulse, respectively. We consider each pulse contributes an instant
spin flip, i.e., $\tau_{p}\rightarrow 0$, in terms of which the evolution operator
$U(\tau_{p})$ can be approximated to
\begin{equation}
U(\tau_{p})\simeq U_{p}=\exp(-\mathrm{i}\frac{\pi}{2}\sigma_{S}^{x}),
\end{equation}
and $\mathcal{T}\simeq2\Delta t$. At time $t=M\mathcal{T}+t^{'}$ with
$M$ the integer part and $t^{'}$ the remainder, the dynamical
evolution is generally dictated by
\begin{equation}
U(t)=\left\{
\begin{array}{ll}
U_{0}(t^{'})[U_{c}(\mathcal{T})]^{M};\quad & t^{'}<\Delta t\\
U_{0}(t^{'}-\Delta t)U_{p}U_{0}(\Delta t)[U_{c}(\mathcal{T})]^{M};\quad & t^{'}>\Delta t
\end{array}%
\right.
\end{equation}
If we focus on small interval $\Delta t$ and stroboscopic time points
$t=M\mathcal{T}$ with $M$ the number of complete pulse cycles, the
dynamical evolution at these time points can be equivalent to
\begin{equation}
U(t)=[U_{\mathrm{eff}}(\mathcal{T})]^{M}=\exp(-\mathrm{i}H_{\mathrm{eff}}t),
\end{equation}
with the effective Hamiltonian~\cite{Hu20123011}
\begin{equation}
H_{\mathrm{eff}}=H_{E}(\bar{h})+J\frac{\varepsilon\mathcal{T}}{4}\sigma_{S}^{z}\otimes\sum_{j=1}^{N}(\sigma_{j,E}^{y}\sigma_{j+1,E}^{x}+\sigma_{j,E}^{x}\sigma_{j+1,E}^{y}),
\end{equation}
where $\bar{h}=h+\varepsilon J/2$. In this case, we obtain the effective
decoherence factor
\begin{eqnarray}
\mathcal{F}_{\mathrm{eff}}(t)&=\prod_{k>0}[1-2n_{x}^{2}\sin^{2}g_{p}+2\mathrm{i}n_{x}\sin^{2}g_{p}\nonumber\\
&\quad\times(n_{y}\cos\theta_{k}-n_{z}\sin\theta_{k})\tanh(2J\beta\Lambda_{k})],
\end{eqnarray}
with
\begin{eqnarray}
n_{x}=\frac{J\varepsilon\sin k/2}{\Lambda_{p}} &, \quad & n_{y}=\frac{\sin k}{\Lambda_{p}},\nonumber\\
n_{z}=\frac{\cos k+\bar{h}}{\Lambda_{p}} &,\quad & g_{p}=2J\Lambda_{p}t,
\end{eqnarray}
where $\Lambda_{p}=\sqrt{(\cos k+\bar{h})^{2}+(1+\varepsilon^{2}\mathcal{T}^{2}/4)\sin^{2}k}$.

The dynamics of negativity of quantumness between $S$ and $A$ with and without pulses control at near zero temperature ($T=10^{-3}$) and at finite bath temperature ($T=5$) is plotted in Fig.~\ref{fig:fig5}(a). It is seen that the negativity of quantumness decays very fast at the critical point of spin bath when without pulses control. By contrast, under pulses control, the negativity of quantumness can be preserved to a quasi-steady value with tiny oscillations. Moreover, the quasi-steady value at finite temperature differs slightly from that at near zero temperature, which means the pulses control is robust against the temperature of environment. To see the effect of frequency of pulses action, we also plot the dynamics of negativity of quantumness under pulses control with different frequencies at specific bath temperature $T=0.5$ in Fig.~\ref{fig:fig5}(b). It is clearly seen that the quasi-steady value of the negativity of quantumness is enhanced and the oscillations are weaken when the frequency of pulses increase (corresponding to the decrease of the period $\mathcal{T}$ for a complete cycle). It is not difficult to conclude that the negativity of quantumness can be kept at its initial value as the frequency increases to a suitable value, which indicates that the quantumness of the system $S$ has been fully protected from decoherence in that situation.

\section{Conclusions}\label{sec:sec5}
In summary, we have proposed a scheme to characterize the non-Markovian dynamics and quantified the non-Markovianity via the non-classicality measured by the negativity of quantumness. As an illustrative model, we employ a dephasing model consisting of a qubit coupled to a thermal Ising spin bath and study the non-classicality of the qubit system by introducing an ancilla. It is shown that revivals of the negativity of quantumnes can be treated as a signature of non-Markovian dynamics. Furthermore, we introduce a measure of the non-Markovianity based on the negativity of quantumness and discuss the influences of bath criticality, bath temperature and bath size on the non-Markovianity. It is found that the non-Markovianity with finite sized spin bath may not be zero at the critical point, which means that the dynamics is not exactly Markovian and  certain revivals of quantumness is still allowed. However, at the critical point, the non-Markovianity is the most sensitive to the bath-size effect. The non-Markovianity decreases in a polynomial manner at non-critical points while it decays in an near exponential manner at the critical point, indicating purely Markovian dynamics in the thermodynamic limit.

Besides, we also find non-trivial behaviours of negativity of quantumness such as sudden change, double sudden changes and keeping constant for different relations between parameters of the initial state. Finally, we apply a series of bang-bang pulses to suppress the decay of non-classicality and identify that the negativity of quantumness can be effectively preserved against the thermal spin bath.

\begin{acknowledgements}
This project was supported by the National Natural Science Foundation of China (Grant No.~11247308 and No.~11274274),
the National Natural Science Foundation of special theoretical physics (Grant No.~11347196),
the natural science foundation of Jiangsu province of China (Grant No.~BK20130162)
and the Fundamental Research Funds for the Central Universities (Grant No.~JUSRP11405).
\end{acknowledgements}



\end{document}